\documentclass[numbers,copyright,creativecommons,runningheads]{eptcs}

\usepackage{breakurl}
\usepackage{amssymb}
\usepackage{graphicx}
\usepackage{multirow}

\usepackage{enumerate}
\usepackage{delarray} 
\usepackage{wrapfig}
\usepackage{url}

\title{Model exploration and analysis for quantitative safety refinement in probabilistic B}
\author{Ukachukwu Ndukwu\thanks{This author acknowledges support from the Australian Commonwealth Endeavor International
Postgraduate Research Scholarship (E-IPRS) Fund.} \ and Annabelle McIver \thanks{This author acknowledges support from the Australian Research Council (ARC) Grant Number DP0879529.}
\institute{Department of Computing, Macquarie University, NSW 2109 Australia.}
\email{\{ukachukwu.ndukwu,annabelle.mciver\}@mq.edu.au}
}
\def\titlerunning{Model exploration and analysis for quantitative safety refinement in probabilistic B}
\def\authorrunning{Ukachukwu Ndukwu \ and Annabelle McIver}

\begin{document}

\newcommand\Lift {\textsf{lift}}
\newcommand\Sem[1] {[\![#1]\!]}
\newcommand\Gets {{:}{=}\,}
\newcommand\Eqn[1] {(\ref{#1})}
\newcommand{\pGCL}{\textit{pGCL~}}
\newcommand\Skip {\textbf{skip}}
\newcommand\Wide[1] {~~~#1~~~}
\newcommand\Ref {\sqsubseteq}
\newcommand\Spot {\mbox{\boldmath$\cdot$}}
\newcommand\Def[1] {Def.~\ref{#1}}
\newcommand\While {\textbf{while}}
\newcommand\Do {\textbf{do}}
\newcommand\Od {\textbf{od}}
\newcommand\Lem[1] {Lem.~\ref{#1}}
\newcommand{\True}{\textbf{true}}
\newcommand{\False}{\textbf{false}}
\newcommand{\Max}{~\sqcup~}

\newcommand{\Inv}{\textit{inv}}

\newcommand{\Prog}{\textit{Prog}}
\def\z@rel#1{\mathrel{\mathstrut{#1}}}
\def\Defs{\ \triangleq \ }
\newcommand{\PC}[1]{\mathbin{\makebox[0em]{~}_{#1}\oplus}}
\newcommand{\Post}{\textit{post}}
\newcommand{\Pre}{\textit{pre}}
\newcommand{\Exp}{\textit{Exp}}
\newcommand{\Pred}{\textit{pred}}
\newcommand{\figg}[1]{Fig.~\ref{#1}}
\newcommand{\tab}[1]{Table~\ref{#1}}
\newcommand{\Wp}{\textit{Wp}}
\newcommand{\Expt}{\textit{Expt}}
\newcommand{\wP}[2]{wp\cdot{#1}\cdot{#2}}
\newcommand{\ccc}[3]{#1\,\,\mbox{$\lhd$}\ #2\ \mbox{$\rhd$}\ #3}
\newcommand{\pp}[1]{\mathbin{\makebox[0em]{~}_{#1}\oplus}}
\newcommand{\IE}{\textrm{i.e.}}
\newcommand{\It}{{\sf it}}
\newcommand{\Ti}{{\sf ti}}
\newcommand{\Min}{~\sqcap~}
\newcommand{\Sch}{\aleph}
\newcommand{\Dist}{{\mathbb D}}
\newcommand{\TSem}[1]{\langle\!|#1|\!\rangle}
\newtheorem{corr}{Corollary}
\newcommand{\Cor}[1]{Cor.\ref{#1}}
\newtheorem{exm}{Example}
\newcommand{\tree}[1]{\mathcal{#1}}
\newcommand{\CompNY}[3]{(#1#2 \Spot #3)}
\newcommand{\EE}[1]{ {\cal E}{#1}}
\newtheorem{Defns}{Definition}
\newtheorem{Lems}{Lemma} 
\newtheorem{Prf}{Proof} 

{\maketitle}

\begin{abstract}
The role played by counterexamples in standard system analysis is well known; but less common is a notion of counterexample in probabilistic systems refinement. In this paper we extend previous work using counterexamples to inductive invariant properties of probabilistic systems, demonstrating how they can be used to extend the technique of bounded model checking-style analysis for the refinement of quantitative safety specifications in the probabilistic B language. In particular, we show how the method can be adapted to cope with refinements incorporating probabilistic loops. Finally, we demonstrate the technique on pB models summarising a one-step refinement of a randomised algorithm for finding the minimum cut of undirected graphs, and that for the dependability analysis of a controller design.\\

\end{abstract}

\setcounter{section}{0}


\section{Introduction}
The B method \cite{BBook} and more recently its successor Event-B \cite{Event-B} comprises a method and its automation for modelling complex software systems. It is based on the top-down refinement where specifications can be elaborated with detail and additional features, whilst the automated prover checks consistency between the refinements. Hoang's probabilistic B or pB \cite{Hoang05} extension  of standard B gave designers the ability to refer to probability and access to the specification of quantitative safety properties.

In probabilistic systems, the generalisation of traditional safety properties allows the specification of random variables whose expected value must always remain above some given threshold. Elsewhere \cite{UNdukwu,YAGA} we have provided automation to check this requirement by analysing pB models using an automatic translation of their quantitative safety specifications as PRISM reward structures \cite{PRISM}. Our technique allows pB modellers to explore the quantitative safety properties encoded within their models to obtain diagnostic feedback in the form of counterexample traces in the case that their model does not satisfy the quantitative specification. Counterexamples become sets of execution traces each with some probability of occurring and jointly implying that the specified threshold is not maintained. Moreover pB's consistency checking enforces inductive invariance of the quantitative safety property, thus the counterexample traces also demonstrate specific points in the models execution where the inductive property fails.

The paradigm of abstraction and refinement supports stepwise development of probabilistic systems aimed at improving probabilistic results. Unfortunately, for quantitative safety specifications (our focus here), a human verifier has no way of inspecting that this requirement is met even though the automated prover readily establishes consistency between the refinements. One way to resolve this uncertainty is to explore algorithmic approaches similar to probabilistic model checking techniques which can provide exact diagnostics summarising the failure (if indeed it exists) of the refinement goal.

In this paper we extend some practical uses of  counterexamples to probabilistic systems refinement with respect to quantitative safety specifications particular to the pB language. We show how to use them to generalise bounded model checking-style analysis for probabilistic programs so that an iteration can be verified by exhaustive search provided that quantitative invariants are inductive for all reachable states. We also show how the use of probabilistic counterexamples in quantitative dependability analysis can be used to determine ``failure modes" and ``critical sets" which thus enables their extension to estimating components severity.

We illustrate the techniques on two case studies: one based on a probabilistic algorithm \cite{Karger} to find the minimum cut set in a graph, and the other a probabilistic design for a controller mechanism \cite{GO10}.

The outline of the paper is as follows.  In Sec.\ref{s2146} we summarise the underlying theory of pB; in Sec.\ref{s1246} we discuss the probabilistic counterexamples we can derive from the models and a bounded model checking approach to probabilistic iteration. In Sec.\ref{csone} we illustrate the technique on the specification of a randomised ``min-cut". We discuss probabilistic diagnostics of dependability in Sec.\ref{s2152} and demonstrate with a case study in Sec.\ref{cstwo}. We discuss related work and then conclude.

\subsubsection{Notation}
Function application is represented by a dot, as in $f.x$ (rather than $f(x)$).
We use  an  abstract finite state space $S$. Given predicate $\Pred$ we write
$\Lift{\Pred}$ for the \emph{characteristic} function mapping states satisfying $\Pred$ to
$1$ and to $0$ otherwise, punning $1$  and $0$ with ``True" and ``False" respectively.\
We write ${\cal E}S$  as the set of real-valued functions from $S$, {\it i.e.} the set of expectations; and
whenever $e, e' \in {\cal E}S$ we write $e \Rrightarrow e'$ to mean that $\mathord{(\forall s\in S. \ e.s \le e'.s)}$.
We let $\mathbb{D}S$ be the set of all discrete probability distributions over $S$; and write $\mathord{Exp.\delta.e = \displaystyle\sum_{s\in S}(\delta.s) \times e.s}$ for the expected value of $e$ over $S$ where $\delta\in\mathbb{D}S$ and $e \in {\cal E}S$. Finally we write $S^{\ast}$ for the finite sequences of states in $S$.

\section{Probabilistic annotations}\label{s2146}

When  probabilistic programs execute they make random updates; in the semantics that behaviour is modelled by discrete probability distributions over possible final values of the program variables.  Given a program $\Prog$ operating over $S$ we write $\Sem{\Prog} : S \rightarrow (S \rightarrow [0,1])$ for the semantic function taking initial states to distributions over final states. For example,
the program fragment

\begin{equation}\label{pInc}
\textit{pInc} \Defs s\Gets s{+}1 \ \PC{p} \ s \Gets s{-}1
\end{equation}
increments state variable $s$  with probability $p$, or decrements it with probability $1{-}p$. The semantics $\Sem{\textit{pInc}}$ for each initial state $s$ is a probability distribution    returning $p$ or  $(1{-}p)$ for (final) states $s' = (s{+}1)$ or $s' = (s{-}1)$ respectively.
Rather   than working with this semantics directly, we shall focus on the dual logical view generalisation of Hoare logic \cite{Hoare}.

Probabilistic Hoare logic \cite{MM04} takes account of the probabilistic judgements that can be made about probabilistic programs, in particular it can express when predicates can  be established  only \emph{with some probability}. However, as we shall see, it is even more general than that, capable of expressing general expected properties of random variables over the program state.
We use \emph{Real}-valued annotations of the program variables interpreted as expectations; a program annotation is said to be valid exactly when the expected value over the post-annotation is at least the value given by the pre-annotation. In detail
\begin{equation}\label{e0446}
\{ \Pre\} ~~ \Prog~~\{ \Post \}~,
\end{equation}
is valid exactly when  $\Exp.\Sem{\Prog}.\Post.s \geq \Pre.s$ for all states $s\in S$, where $\Post$ is interpreted as a random variable over final states and $\Pre$ as a real-valued function.

With our notational convention, a correct annotation for \textit{pInc} (at \Eqn{pInc}) is given by the triple
\begin{equation}\label{e0420}
\{ p \times \Lift{(s=-1)} + (1{-}p) \times \Lift{(s=1)} \} ~~~~ \textit{pInc}~~~~\{ \Lift{(s = 0)}\}~,
\end{equation}
which expresses the probability of establishing the state $s=0$ finally, depending on the initial state from which \textit{pInc} executes. Thus if the initial state is
 $s=-1$ then that probability is $p$, but it is  $(1{-}p)$ if the initial state is  $s=1$.

Rather than use the distribution-centered semantics outlined above, we shall use a generalisation of  Dijkstra's weakest precondition or $\Wp$ semantics defined on the program syntax of the probabilistic Guarded Command Language or \pGCL \cite{MM04}. The semantics of the language is set out in  \figg{fig:semantics}.  As for standard $\Wp$ this formulation allows annotations to be checked mechanically \cite{Hoang05,Hurd}; moreover
we see that annotation \Eqn{e0446} is valid exactly when $\Pre \Rrightarrow \Wp.{\Prog}.{\Post}$.

\begin{figure}
\begin{center}
\[
\small
\begin{array}{lll}
Name &\Prog & \Wp.{\Prog}.{\Expt}\\
\hline\\
\mbox{identity} & \mbox{\Skip} &  \mbox{$\Expt$}   \\
\mbox{assignment} & \mbox{$x := f$}  & \mbox{$\Expt[x := f]$}\\
\mbox{composition} & \mbox{{$Prog;Prog'$}} & \mbox{$\Wp.{Prog}.{(\wP{Prog'}{\Expt})}$}\\
\mbox{choice} & \mbox{$\ccc{Prog}{G}{Prog'}$}& \mbox{$\ccc{\Wp.{Prog}.{\Expt}}{G}{\Wp.{Prog'}.{\Expt}}$} \\
\mbox{probability} & \mbox{$Prog \pp{p} \ Prog'$} & \mbox{$\Wp.{Prog}.{\Expt} \ \pp{p} \ \Wp.{Prog'}.{\Expt}$}\\
\mbox{nondeterminism} & \mbox{${Prog \ \Min \ Prog'}$} & \mbox{$\Wp.{Prog}.{\Expt} \ $min$ \ \Wp.{Prog'}.{\Expt}$}\\
\mbox{weak iteration} & \mbox{\It \ $Prog$ \ \Ti} & \mbox{$\nu X \bullet (\Wp.{Prog}.{X}\ $min$ \ \Expt)$}\\
\\
\hline\\

\end{array}
\]
\footnotesize{Given a program command $Prog$ and expectation $\Expt$ of type ${\cal E} S$, $\textit{Wp}.\Prog$ is of type ${\cal E} S \rightarrow {\cal E} S$. Note also that we write $\Exp.(\Sem{\Prog}.s).\Expt$} to mean $\textit{Wp}.\Prog.\Expt.s.$
\end{center}
\caption{\rm Structural definition of the expectation transformer-style semantics.} \label{fig:semantics}
\end{figure}

In this paper we shall concentrate on certifying probabilistic safety expressible using probabilistic annotations.
  Informally, a probabilistic safety property is a random variable whose expected value cannot be decreased on execution of the program. (This idea generalises standard safety, where the \emph{truth} of a safety predicate cannot be violated on execution of the program.) Safety properties are characterised by \emph{inductive invariants}: for example the valid annotation $\{\Expt {\times} \Lift{\Pred} \} ~ \Prog ~ \{\Expt\}$ says that  $\Expt$ is an inductive invariant for $\Prog$ provided it is executed in an initial state satisfying $\Pred$.
To illustrate, the annotation
 \begin{equation}\label{e1325}
\{ s \} ~~~~ \textit{pInc}~~~~\{  s\}~,
\end{equation}
 means that the expected value of $s$ is never decreased (and it is therefore only valid if $p \geq 1/2$).

Inductive invariants will be  a  significant component of the refinement of quantitative safety specifications in our pB machines, to which we now turn.

\subsection{Probabilistic safety and refinement in pB}

Probabilistic B or pB \cite{Hoang05}, is an extension of standard B \cite{BBook} to support the specification and refinement of probabilistic systems.
Systems are specified by a collection of \emph{pB machines} which consist of
operations describing possible program executions, together with variable declarations and invariants prescribing correct behaviour.

The machine set out in \figg{fig:DETMachine} illustrates some key features of the language. There are two operations --{\it OpX} and  {\it OpY}-- which can update a variable $cc$. {\it OpX} can either increment {\it cc} by 1 or decrement it by the same value with probability $p$ or $(1-p)$ respectively, while {\it OpY} just resets the current value of {\it cc} to 0. In general, operations can execute only if their preconditions hold. But in the absence of preconditions as in this case, the choice of which operation to execute is made nondeterministically.

The remaining clauses ascribe more information to the variables, constants and behaviour of the operations. Declarations are made in the CONSTANTS and VARIABLES clauses; PROPERTIES and SEES clauses state assumed properties and context of the constants and variables. The INVARIANT clause sets out invariant properties. The expression in the INITIALISATION clause must establish the invariant and the operations {\it OpX} and {\it OpY} must maintain it afterwards.

We shall concentrate on the EXPECTATIONS clause\footnote{However, Hoang \cite{Hoang05} showed that another way to check that a real-value $\Omega$ is indeed an expectation is to evaluate the language-specific boolean function $expectation(\Omega)$.\ Therefore we shall interchangeably use both forms to denote expectations-based expressions with no loss of generality.}, which was introduced by Hoang \cite{Hoang05} to express quantitative invariant or safety properties. The form of an EXPECTATIONS clause is given by

\begin{figure}
\begin{center}
\scriptsize{
\begin{tabular}{ll}
\hline\\
\mbox{{\bf MACHINE}} & \mbox{Faulty} \\
\mbox{{\bf SEES}} & \mbox{Int\_TYPE, Real\_TYPE}\\
\mbox {{\bf CONSTANTS}} & \mbox{$p$}\\
\mbox {{\bf PROPERTIES}} & \mbox{$ p\in REAL \wedge p \ge real (0) \wedge p \le real (1)$}\\
\mbox {{\bf VARIABLES}} & \mbox{$cc$}\\
\mbox{{\bf INVARIANT}} & \mbox{$cc \in\mathbb{N}$} \\
\mbox{{\bf INITIALISATION}} & \mbox{$cc := 0$} \\
\mbox{{\bf OPERATIONS}} &\\
& \mbox{\quad  \ \ OpX $\Defs$ {\bf BEGIN}} \\
& \qquad \qquad \qquad \mbox{{\bf \ PCHOICE} \ $p$ \ {\bf OF} \ $cc : = cc + 1$ } \\
& \qquad \qquad \qquad \mbox{{\bf \ OR} \ $cc : = cc - 1$ \ {\bf END;}} \\
& \quad \mbox{{  \ OpY} $\Defs$ $cc := 0$} \\
\cline{1-2}
\multicolumn{1}{|l}{\mbox{{\bf EXPECTATIONS}}}  & \multicolumn{1}{l|}{\mbox{$real (0) \ \Rrightarrow \ cc$}}\\\cline{1-2}\\
\mbox{{\bf END}} &\\
\hline\\
\end{tabular}
\noindent
}\\
{\footnotesize Bold texts on the left column capture the fields (or clauses) used to describe the machine. The {\bf PCHOICE} keyword introduces a probabilistic binary operator; the {\bf EXPECTATIONS} clause expresses the notion of probabilistic quantitative safety.}
\caption{\rm A simple pB machine.}\label{fig:DETMachine}
\end{center}
\end{figure}

\begin{equation}\label{e1352}
E \Rrightarrow \Expt~,
\end{equation}
where both $E$ and $\Expt$ are expectations. It specifies that the expected value of $\Expt$  should always be \emph{at least} $E$, where the expected value is determined by  the  distribution over the state space after any valid execution of the machine's operations, following its initialisation. Hoang showed that this is guaranteed  by the following valid annotations:

\begin{equation}\label{e1137}
\{ E \} \Wide{\textit{init}} \{\Expt \}   ~~~\quad \textit{and} ~~~ \quad\{ \Lift{\Pred} {\times}\Expt \} \Wide{\textit{Op}} \{\Expt \}~,
\end{equation}
where   \textit{Op} is any operation with precondition $\Pred$ and \textit{init} is the machine's initialisation.  In what follows we shall refer to \Eqn{e1137} as the \emph{proof obligations} for the associated expectations clause \Eqn{e1352}.

 Checking the validity of program annotation, and in particular inductive invariants for loop-free program fragments can be done mechanically based on the semantics set out in \figg{fig:semantics}.  In some cases the proof obligation cannot be discharged, and there are two possible reasons for this. The first possibility is that $\Expt$ is too weak to be an inductive invariant for the machine's operations, and must be strengthened by finding $\Expt' \Rrightarrow \Expt$ so that the original safety property can be validated. The second possibility is that the machine's operations actually violate the probabilistic safety property.

The same reasoning can be extended to refinement of abstract pB machines.\ We note that quantitative safety specifications in pB can also be refined in the usual way with respect to expectation pairs. Thus another way of expressing (\ref{e1352}) is to say that any program command $P$ satisfies the bounded expectation pair $[E, \Expt]$ if execution from its initial state guarantees that
\begin{equation}\label{ref}
E \Rrightarrow \Wp.{P}.\Expt.
\end{equation}
Refinement is then implied by the ordering of program commands so that more refined programs improve probabilistic results. More specifically, we write
\begin{equation}\label{ref2}
\mathord{P \Ref Q \Wide{\textit{iff}} \CompNY{\forall}{E \in \EE{S}}{\Wp.P.E \Rrightarrow \Wp.Q.E}},
\end{equation}
to mean that the program command $Q$ is a refinement of the program command $P$. In addition we note that the preservation of an expression like (\ref{e1352}) is implied by the {\it monotone} property of \Wp.

The refinement of abstract pB machines embedding quantitative safety statements is dealt with in the language framework by introducing the IMPLEMENTATION and REFINES clauses. The former clause specifies the refinement of an abstract machine specified in the latter clause. The refinement process is then aimed at preserving the bounds of expectations in the original specification statement (the machine to be refined) so that the validity of an expression like (\ref{e1137}) can be checked mechanically.

Our aim in the next section is to use probabilistic counterexamples adopted in model checking techniques to interpret failure of proofs of refinement of probabilistic machines in the pB language. We will find that a counterexample is a trace (or a set of traces) from the initialisation to a state where the inductive invariant fails to hold after inspecting the EXPECTATIONS clause over the refinement.


\section{Probabilistic safety in Markov Decision Processes}\label{s1246}

In abstract terms $\pGCL$ programs and pB machines may be modelled  as a Markov Decision Process (\textit{MDP}).  Recall that an \textit{MDP} combines the notion of probabilistic updates together with some arbitrary choice between those updates \cite{MDP}: that combination of probabilistic choices together with nondeterministic choices is  present in $\pGCL$ and captures both features.

In this section we summarise pB models\footnote{We note that an abstract pB model begins with the MACHINE keyword while a refinement is a pB model that begins with the IMPLEMENTATION keyword.} and their quantitative safety specifications in terms of \textit{MDP}s, and show how to apply model checking's search techniques for counterexamples to prove quantitative safety as a first step towards generalising standard bounded model checking verification. Inductive invariance is then crucial to the application of exhaustive state exploration for the intended goal.

Here we consider an \textit{MDP} expressed as a nondeterministic selection $\mathord{P \Defs P_0 \Min \dots \Min P_n}$ of deterministic $\pGCL$ programs, where the nondeterminism corresponds to the arbitrary choice, and each $P_i$ corresponds to the probabilistic update for a choice  $i$.  When $P$ is iterated for some arbitrarily-many steps, we identify a \emph{computation path} as a finite sequence of states $\langle s_0, s_1, s_2, \dots, s_n\rangle$ where each $(s_i, s_{i+1})$ is a probabilistic transition of $P$, {\it i.e.}  $s_{i+1}$ can occur with non-zero probability by executing $P$ from $s_i$.  Note that the choice (between $0 \dots n$) can depend on the previous computation path since for example guards for the individual operations $P_i$ must hold for their selection to be enabled.

Standard safety properties identify a set of ``safe" states --- the safety property then holds provided that all states reachable from the initial state under specified state transitions are amongst the selected safe states. A generalisation of this for probabilistic systems specifies thresholds on the probability for which the reachable states are always amongst the safe states. The quantitative safety properties encapsulated by the EXPECTATIONS clause are even more general than that, allowing the possibility to specify thresholds on arbitrary expected properties.  The next definition sets out the mathematical model for interpreting general quantitative safety properties.

Since \textit{MDP}s contain both nondeterministic and probabilistic choice, taking expected values only makes sense over well-defined probability distributions --- we need to resolve the nondeterministic choice in all possible ways to yield a set of probability distributions. The next definition sets out a mechanism for doing just that.

\begin{Defns}\label{d0638}
Given a program $P$, an \emph{execution schedule} is a map $ \Sch: S^* \rightarrow \Dist$S so that $\Sch.\alpha \in \Sem{P}.s$ picks a particular resolution of the nondeterminism in $P$ to execute after the  trace $\alpha$, where $s$ is the last item of $\alpha$. (A more uniform formalisation would give the distribution of initial states as $ \Sch.\langle\rangle$; but we prefer to give initial states explicitly.)%
\end{Defns}
%
Once a particular schedule has been selected, the resulting behaviour generates a probability distribution over computation path. We call such a distribution a \emph{probabilistic computation tree};  such distributions are well-defined  with respect to Borel algebras based on the traces. 
\begin{Defns}\label{d0354}
Given a program $P$, initial state $s_0$ and execution schedule $ \Sch$, we define the corresponding trace distribution $\TSem{P_ \Sch}.s_0$ of type $S^* \rightarrow [0,1]$ to be
%
\[
\begin{array}{lrcl}
& \TSem{P_ \Sch}.s_0.(s')
&\Defs
&1~\textrm{if}~ s'=s_0~\textrm{else}~0 \\
\textrm {and}
&\TSem{P_ \Sch}.s_0.(\alpha ss')
&\Defs
&\TSem{P_ \Sch}.s_0.(\alpha s){\times}{ \Sch.(\alpha s)}.s'
\end{array}
\]
\end{Defns}
Computation trees of finite depth generate a  \emph{distribution over endpoints} as follows.
If we take $K$ steps from some initial $s_0$ according to the schedule $\Sch$, then the probability of ending in state $s'$ is given by
\[
\Sem{P^K_ \Sch}.s_0.s' \Defs \sum_{|\alpha|=K} \TSem{P_ \Sch}.s_0.(\alpha s')~.
\]

General quantitative safety properties are intuitively specified via a numeric threshold $e$ and a random variable $\Expt$ over the state space $S$: the expected value of $\Expt$  with respect to any distribution over endpoints should never fall below the threshold $e$.

\begin{Defns}\label{d1717}
Given threshold $e$ and an expectation $\Expt$ the \emph{general quantitative safety} property is satisfied by the program $P$ if for all schedules $\Sch$ and $K \geq 0$, we have that $\Exp.\Sem{P^K_ \Sch}.\Expt.s_0 \geq e$.
\end{Defns}

The probabilistic Computation Tree Logic or \textit{pCTL} \cite{pCTL} safety property, which places a threshold on the probability that the reachable states always satisfy the identified ``safe" states is expressible using \Def{d1717} via  characteristic expectation $\Lift{safe}$. However many more general properties are also expressible, including expected time complexity \cite{PRISM}.

We shall be interested in identifying situations where the inequality in \Def{d1717} does not hold. Evidence for the failure is a (finite)
computation tree whose distribution over endpoints illustrates the failure to meet the threshold.

\begin{Defns}\label{d0719}
Given a probabilistic safety property, a {failure tree} is defined by a scheduler $\Sch$ and an integer $K \geq 0$ such that $\Exp.\Sem{P^K_ \Sch}.\Expt.s_0 < e$.
\end{Defns}

Elsewhere \cite{Ndukwu11} we showed that if  $\Expt$ is an inductive invariant, then the safety property based on $\Expt$ is implied, provided that $e \leq \Expt.s_0$. In fact, given a failure tree, there must be some finite trace $\alpha$ such that
$ \TSem{P_ \Sch}.s_0.(\alpha s) > 0$ and $\Wp.(P\Min \Skip).\Expt.s < \Expt.s$ \cite{Ndukwu11}.
Thus, as for standard model checking, we are able to locate specific traces which lead to the failure of the invariant property.  We define a counterexample to \emph{inductive invariance} as follows.

\begin{Defns}\label{d1747}
Given a scheduler $\Sch$, an expectation  $\Expt$ and a program $P$, a {counterexample to inductive invariance}  safety property is a trace $(\alpha s)$ which can occur with non-zero probability, and such that $\Wp.P.\Expt.s < \Expt.s$. A state such as s is a witness to failure.
\end{Defns}

But note that in practice there will be a number of counterexamples. Our technique is able to identify them all given any depth $K$ of computation. Next we discuss how the strategy can be extended to probabilistic loops reasoning.

\subsection{Analysis of loops}\label{s2150}
We assume a loop of the form $
\textit{loop} \Defs \While \ G \ \Do  \ body \ \Od \,
$
where $G$ is a predicate over the program state representing the loop guard; $body$ is a probabilistic program consisting of a finite nondeterministic choice over probabilistic updates. Our aim in this section is to generalise the technique of bounded model checking to prove the safety assertion of the form
\begin{equation}\label{l0926}
\{e \} ~~\textit{loop} ~~ \{\Inv\} ~.
\end{equation}

In the case that \Eqn{l0926} does not hold there must be a failure tree (\Def{d0719}) to witness that fact, together with a set of failures to inductive invariance of $\Inv$. We shall be interested in the complementary problem, in the case that the property does hold. For  standard programs this can be established by exhaustively searching the reachable states; any revisiting of a state terminates the search at that point, so that the method is complete for finite state programs: either a counterexample is discovered or all reachable states are visited, and each one checked for satisfaction of the (qualitative) safety property.

The situation is not quite so straightforward for probabilistic programs, and that is because the technique of exhaustive search does not generalise immediately to quantitative safety properties. However {\it via} inductive invariants it does. Consider the program which repeatedly sets a variable $x$ uniformly in the set $\{0, 1, 2\}$ after the initialisation $x:= 1$, and terminates whenever $x$ is set to $2$. In this case we might like to verify the safety property that $x \in \{1, 2\}$ with probability at least $1/2$. Expressed as an assertion, it becomes

\begin{equation}\label{l0943}
\{1/2\} ~~~~~~~ \begin{array}{l}  x:= 1; \While ~ (x = 1) ~\Do ~~~ x\Gets 0 \PC{1/3}( x \Gets 1 \PC{1/2}  x \Gets 2)~\Od \end{array}~~ ~~~~\{\Post\}~,
\end{equation}
where $\Post \Defs \{ \Lift{(x \in \{1, 2\})}\}$. A \emph{quantitative inductive invariant}  establishing that fact is given by $x/2$, expressing the probability that the safety property is always satisfied at that state. (When $x$ is $2$ that probability is $1$, when $x$ is $1$, it is $1/2$ and when $x$ is $0$ it is $0$.) In fact the property \Eqn{l0943}  is equivalently formulated  by setting $\Post \Defs x/2$, which can be seen as a strengthening of $\{ \Lift{(x \in \{1, 2\})}\}$.

Since the triple (\ref{l0943}) does indeed hold, no failure trees exist; more generally, in standard model checking and for finite state spaces such a failure to establish the presence of a failure tree can be converted to a proof that the property holds (provided all reachable states are examined).
For probabilistic systems however, it is not clear when to terminate a state exploration, since  $\Exp.\Sem{\textit{body}^K_ \Sch}.x/2$ steadily approaches $1/2$ from above (where here $\textit{body}$ is taken to be the guarded loop body of \Eqn{l0943}).  However we can recover the termination property even for probabilistic systems by looking at inductive invariants, as the next lemma shows.

\begin{Lems}\label{l0948}
Let $P$ be a probabilistic program operating over a finite state space $S$; let $s_0$ be the initial state. If for all states $s$, reachable from $s_0$ under executions via $P$, the inductive invariance property $\Wp.P.\Inv.s \geq \Inv.s$ holds, then $\Exp.\Sem{P^K_ \Sch}.\Inv \geq \Inv.s_0$ for all $K$ and schedules $\Sch$.
\begin{Prf} (Sketch)
We use proof by induction on $K$.

When $K=1$ we note that $\Exp.\Sem{P^1_ \Sch}.\Inv \geq \Inv.s_0$ is a consequence of the assumption since $\Exp.\Sem{P^1_ \Sch}.\Inv \geq  \Wp.P.\Inv.s_0$.

For the general step, we observe similarly that $\Exp.\Sem{P^{K+1}_ \Sch}.\Inv \geq \Exp.\Sem{P^K_ \Sch}.(\Wp.P.\Inv)$. The result follows through monotonicity of the expectation operator.
\end{Prf}
\end{Lems}

\Lem{l0948} implies that we can use exhaustive search to verify quantitative safety properties using inductive invariants and exhaustive state exploration. The search terminates once all reachable states have been verified as satisfying the inductive property. In the case of \Eqn{l0943}, using $x/2$  for the invariant, each of the three states satisfies the inductive property. Next we summarise a prototype tool framework for locating and presenting counterexamples.

\section{Automating counterexamples generation}
YAGA \cite{YAGA} is a prototype suite of programs for inspecting safety specifications of abstract pB machines and their refinements. Importantly, it allows a pB machine designer to explore experimentally the  details of system construction in order to ascertain the cause(s) of failure of a pB safety encoding as in (\ref{e1352}).

YAGA inputs a pB machine or its refinement violating a specific safety property expressed in its EXPECTATIONS clause, and generates its equivalent MDP representation in the PRISM language \cite{PRISM}. PRISM is a probabilistic model checker that permits pB models as MDPs in the tool framework and thus can investigate critical expected values of random variables as ``reward structures'' --- a part of PRISM's specification language. PRISM can then be used to explore the computation of $\Exp.\Sem{P_ \Sch^K}.\Expt.s_0$ for values of $K\geq 0$, and thus (modulo computing resources) can determine values of $K$ for which the expectations clause fails. If such a $K$ is discovered, YAGA is able to extract the resultant failure tree as an ``extremal scheduler'' that fails the inductivity test. The extremal scheduler is a transition probability matrix which gives a description of the best (or worst-case) deterministic scheduler of the PRISM representation of an abstract `faulty' pB machine --- {\it i.e.} one whose probability (or reward) of reaching a state where our intended safety specification is violated is maximal (or minimal).

Finally, YAGA analyses the  resultant extremal scheduler using algorithmic techniques set out in \cite{Ndukwu11} and generates `the most useful' diagnostic information composed of finite execution traces as sequences of operations and their state valuations leading from the initial state of the pB machine to a state where the property is violated. Details of the underlying theory of YAGA, its algorithms and implementation can be found elsewhere \cite{YAGA,Ndukwu11}. In the next section we discuss practical details on how to use exhaustive search of pB machines to verify compliance of  inductivity for finite probabilistic models.

\section{Case study one: min-cut}\label{csone}
We discuss one of Hoang's  pB models  \cite{Hoang05}: a randomised solution to finding the ``minimum cut" in an undirected graph. The probabilistic algorithm is originally due to Karger \cite{Karger}. We also report experimental results after running our diagnostic tool.

Let an undirected graph be given by  $(N, E)$ where $N$ is a set of nodes and $E$ is a set of edges. The graph is said to be \emph{disconnected} if $N$ is a disjoint union of two nonempty sets $N_0, N_1$ such that any edge in $E$ connects nodes in $N_0$ or $N_1$; a graph is \emph{connected} if it is not disconnected.  A \emph{cut} in a connected graph is a subset $E' \subseteq E$ such that $(N, E\backslash E')$ is disconnected; a cut is minimal if there is no cut with strictly smaller size. Cuts are useful in optimisation problems but are difficult to find. Karger's algorithm uses a randomisation technique which is not guaranteed to find the minimal cut, but only with some probability. 
\begin{figure}
\begin{center}
\scriptsize{
$
\begin{array}{ll}
 \hline\\
\noindent
\mbox{{\bf IMPLEMENTATION}} & \mbox{contractionImp} \\
\mbox{{\bf REFINES}} & \mbox{contraction} \\
\mbox{{\bf SEES}} & \mbox{Bool\_Type, Int\_TYPE, Real\_TYPE}\\
\mbox {\bf OPERATIONS} &\\
\quad \quad \mbox {ans $\longleftarrow$ {\bf contraction} ($NN$) $\Defs$ } &  {\bf VAR} \ $nn$ \ {\bf IN}\\
\quad \quad\quad \quad \mbox{{\bf }} & \mbox{$nn := NN; ans := TRUE;$} \\\\
& \mbox {{\bf WHILE} (nn $>$ 2) {\bf DO}}\\
&  \mbox {$ans \longleftarrow {\bf merge}(nn, ans)$;}\\
&  \mbox {$nn := nn - 1$ }\\
& \mbox{{\bf VARIANT} {\quad $nn$}}\\
& \mbox{{\bf INVARIANT} {$nn \in \mathbb{N} \wedge nn \le NN \wedge 2 \le nn \wedge ans \in BOOL \ \wedge$}}\\
& \mbox{$expectation( frac (2, nn \times (nn - 1 )) \times \Lift{ans})$}\\
\mbox{{\bf END;}}\\
\mbox{{\bf END}} &\\
\hline
\end{array}
$
\noindent
}
{\footnotesize}.
\caption{\rm A pB refinement of the contraction specification of the Mincut algorithm.}\label{fig:mincut}
\end{center}
\end{figure}
The idea of the algorithm is to use a ``contraction" step, where first an edge $e$ connecting two nodes $(n_1, n_2)$ is selected at random and then a new graph created from the old by ``merging" $n_1$ and $n_2$ into a single node $n_{12}$; edges in the merged graph are the same as in the original graph except for edges that connected either $n_1$ or $n_2$. In that case if $(n_1, a)$, say was an edge in the original graph then $(n_{12}, a)$ is an edge in the merged graph. We keep merging while the number of nodes is greater than 2. The specification of the merge function for an initial number of nodes $NN$ is such that
\[
ans \longleftarrow {\bf merge} (nn, aa) \Defs  nn \in NN \wedge aa \in BOOL \ | \ ans := ({\bf false} \ _{\le \PC{2/nn}} \ aa).
\]
It expresses that with a probability of at most $2/nn$, the minimum cut will be destroyed by the contraction step. Otherwise the minimum cut is guaranteed to be found. Contraction satisfies an interesting combinatorial property which is that if the edge is chosen uniformly at random from the set of edges then the merged graph has the same minimum cut as does the unmerged graph with probability at least $2/(NN(NN{-}1))$. Although this probability can be small, it can be amplified by repeating the algorithm to give a probability of assurance to within any specified threshold.

The pB implementation in \figg{fig:mincut} sets out part of the refinement step for the min-cut algorithm. The refinement describes an iteration where the {\bf merge} function is called to perform the contraction described above. The result of a call to merge is that the number of nodes in the graph (given by the variable $nn$) is diminished by $1$ and either the original minimum cut is preserved (with probability mentioned above), or it is not; the Boolean {\it ans} is used to indicate which of these possibilities has been selected.

Here we use the $expectation(.)$ function to check that the expression $\Lift{ans} \times 2/(nn(nn{-}1))$ simplifies to an inductive property; that is, that the probability of preserving the minimum cut should always be at least $2/(nn(nn{-}1))$ while $ans$ remains $\True$, but is $0$ if $ans$ ever becomes $\False$. Note that if this property holds then we are able to deduce exactly that the overall probability that the original minimum cut is preserved when the graph is merged to one of $2$ nodes is the theoretically predicted  $2/(NN(NN{-}1))$.

Next we describe bounded model checking style experiments to analyse the refinement.


\subsection{Experiments for min cut}

\subsubsection{Counterexample diagnostics}
\begin{figure}
\begin{center}
\includegraphics[scale = 0.5]{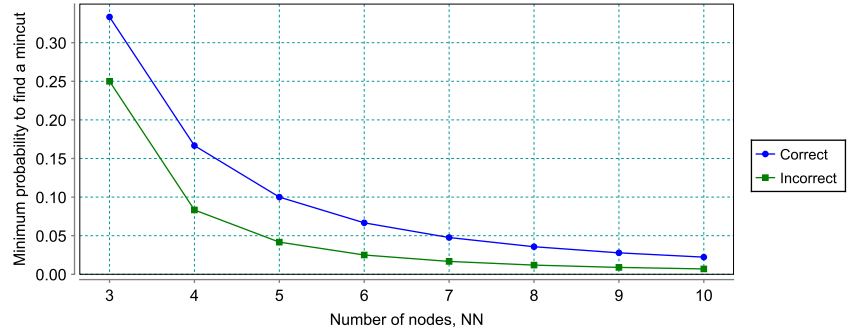}
\end{center}
\caption{Graph comparing the probabilities to find a min-cut for the correct and incorrect implementations of the contraction specification of the mincut algorithm. The incorrect implementation is where we have introduced a high probability in the left branch of the {\bf merge} operation thus forcing the variable {\it ans} to become {\bf false} often. }\label{fig:mincuterror}
\end{figure}

In our first experiment we introduce an error \footnote{We set the probability of choosing the left branch in the merge specification to be ``at most'' 3/4 so that the new specification becomes  $ans := ({\bf false} \  _{\le\PC{3/4}} \ aa)$} in the design of the {\bf merge} function. The graph depicted in \figg{fig:mincuterror} shows a failure to preserve the expected probability threshold of the mincut algorithm. Specifically the graph shows that the probability falls below $2/(NN(NN{-}1))$. An examination of the resultant failure tree produces the counterexample depicted in \figg{fig:trace}. It clearly reveals a problem ultimately leading to a witness after executing the {\bf merge} operation.

\begin{figure}
\begin{center}
\small
     \begin{verbatim}
    ******* Starting Error Reporting  for Failure Traces located on step 2 *********
                Sequence of operations leading to bad state ::>>>
                            [{INIT} (3,true), {Skip} (3,true)
                    Probability mass of failure trace is:>>>> 1
                ************ Finished Error Reporting***************
\end{verbatim}
\caption{\footnotesize Diagnostics detailing a failure of the inductive invariance at the implementation step (for $NN = 3$) involving the {\bf merge} operation. Note that this is a counterexample since the execution of the merge operation will result in an endpoint distribution which yields a decreased expectation (see Def.\ref{d1747}). That is, there is a witness $s$ ($nn = 3$, $ans = true$) such that $\Wp.{\bf merge}.2/(nn(nn - 1)).s = 1/12 < 2/(nn(nn - 1)).s = 1/3$. Note that every trace component of the counterexample is marked with a pair which denotes the state valuations of the program variables occurring in the EXPECTATIONS clause, in this case ($nn$, $ans$).}\label{fig:trace}
 \end{center}
\end{figure}

\subsubsection{Proof of correctness for small models}
In the next experiment we fix the error in the {\bf merge} function and attempt a verification of mincut for specific (small) model sizes. In particular, we use YAGA to check that the EXPECTATIONS clause satisfies the inductive property for all reachable states. The result is shown in \tab{f1535}. It depicts the various sizes of the PRISM model relative to the number of nodes $NN$ of interest of the original graph.

\begin{table}
\begin{center}
\small{
\begin{tabular}{|c|c|c|c|}
 \hline
 \multicolumn{4}{|c|}{PRISM model checking results for mincut algorithm for varying node sizes}\\
 \hline
{\bf NN} & {\bf States, transitions}  & {\bf Probability to find a mincut} & {\bf Duration (secs)} \\ \hline
10 & 72517, 128078 & 2.2222 E-1 & 18.046 \\ \hline
50 & 412797, 732718 & 8.1633 E-4 & 131.363 \\ \hline
100 & 797647, 1416518 & 2.0202 E-4 & 277.605 \\ \hline
\hline
\end{tabular}
}
\caption{\rm Performance result of inductive invariance checking for mincut}\label{f1535}
\end{center}
\end{table}

\section{Probabilistic diagnostics of dependability}\label{s2152}
In this section we investigate how the use of probabilistic counterexamples can play a role in the analysis of dependability, especially in compiling quantitative diagnostics related to specific ``failure modes".

We assume a probabilistic model of a critical system, and we shall use the notation and conventions set up in Sec.\ref{s1246}. In addition, we shall reserve the symbol $F$ for a special designated state  corresponding to ``complete failure"; in the case that a system completely fails (\IE\ enters the $F$ state) we shall posit that no more actions are possible. In the design of dependable systems, one of the goals is to understand what behaviours lead to complete failure, and how the design is able to cope overall with the situation where partial failures occur. For example, the design of the system should be able to prevent complete failure even if one or more components fail. Regrettably, some combinations of component failures will  eventually lead to complete failure --- those combinations are usually referred to as {\it failure modes}. In such cases, dependability analysis would  seek to confirm that the relevant failure modes were very unlikely to occur and also, to produce some estimate of the time to complete failure once the failure mode arose.

We first set out definitions of failure modes and related concepts relative to an MDP model.  In the definitions below we refer to  $P$ as an MDP, with  $F$ a designated state to indicate ``complete failure", such that the annotation $\{ F\} ~P ~\{F\}$ holds.
Let $\phi$ be a predicate over the state space and $\alpha$ a sequence of states indicating an execution trace of $P$. We define the the path formula  $\diamond \phi$  to be  $(\diamond \phi).\alpha = \True$ if and only if there is some $n \geq 0$ such that $\alpha.n$ satisfies $\phi$, corresponding to the usual definition of ``eventuality" \cite{pCTL}.

Our next definition identifies a failure mode: it is a predicate which, if ever satisfied, leads to failure with probability $1$.  We formalise this as the \emph{conditional probability} {\it i.e.} that $F$ occurs given that the failure mode occurs.  We use the standard formulation for conditional probability: if $\mu$ is a distribution over an event space, we write $\mu.A$ for the probability that event $A$ occurs and $\mu.(A \ | \ B)$ for the probability that event $A$ occurs given that event $B$ occurs. It is defined by the quotient $\mu.(A \land B)/ \mu.B$.

Standard approaches for dependability analysis largely rely on the failure mode and effects analysis or (FMEA) \cite{FMEA} for identifying a ``critical set" ---  the minimal set of components whose simultaneous failure constitutes a failure mode. Next we shall show how probabilistic model checking can be used to generalize this procedure.
\begin{Defns}\label{d1300}
Let $P$ be an MDP and let $\Sch$ be a scheduler; we say that a predicate $\phi$ over the state space is a {failure mode} for $\Sch$  if the probability that $F$ occurs given that $\phi$ ever holds is $1$:
\[
\Sem{P^K_ \Sch}.s_0.(\diamond F \ |\ \diamond \phi) \Wide{=} 1~,
\]
where we write $\Exp.\Sem{P^K_ \Sch}.s_0.(\diamond F \ | \ \diamond \phi)$ as the conditional probability over traces such that $F$ is reachable from the initial state $s_0$ given that $\phi$ previously occurred. We say that $\phi$ defines a {critical set} if $\phi$ is a weakest predicate which is also a failure mode.
\end{Defns}

Given the assumption that once the system enters the state $F$, it can never leave it, \Def{d1300} consequently identify states of the system which certainly lead to failure.

Once a critical set has been identified, we can use probabilistic analysis to give  detailed quantitative profiles, including the probability that it occurs, and estimates of the time to complete failure once it has been entered. The probability that a critical set $\phi$ occurs for a scheduler $\Sch$ is given by $\Exp.\TSem{P_\Sch}.(\diamond \phi)$. The next definition sets out the basic definition for measuring the time to failure --- it is based on the conditional probability measured at various depths of the execution tree.

\begin{Defns}\label{d1343}
Let $P$ be an MDP, $\Sch$ a scheduler and let $K$ refer to the depth of the associated execution tree. Furthermore let $\phi$ be a critical set. The probability that complete failure has occurred at depth $K$ given that $\phi$ has occurred is given by:
\[
\Sem{P^K_ \Sch}.s_0.(\diamond F \ | \ \diamond \phi)~.
\]
\end{Defns}
Thus even though a failure mode has been entered, the analysis can determine the approximate depth of computation $k \le K$ before complete failure occurs.

\subsection{Instrumenting model checking with failure mode analysis}

In this section we describe how the definitions above can be realised within a probabilistic model checking environment in order to identify and analyse particular combinations of actions that lead to failure.\footnote{Note that YAGA computes probabilities over endpoints rather than over traces, thus we assume that failure modes can be identified by entering a state which persists according to \Def{d1300}. These will be deadlock states of the MDP being analysed.}

\subsubsection{\bf Identification of failure modes} The first task is to interpret \Def{d1300} as a model checking problem: this relies on the calculation of \emph{conditional probabilities} which is not usually possible using standard techniques. However, adopting the more general expectations approach --- instrumented as reward structures of MDPs --- we are able to compute lower bounds on conditional probabilities after all.

\begin{Lems}\label{l1047}
Let $P$ be a $\pGCL$ program and $\Sch$ a scheduler, $X,C$ are predicates over $S$, and $\lambda$ is a real value  at least $0$. Starting from an initial state $s_0$, the following relationship holds.%
\footnote{This expression may be generalised to allow for non-determinism: $\Exp.\Sem{P}.s_0.(\Lift{(C \land X)} - \lambda {\times}\Lift{C}) \geq 0 \Wide{\textit{iff}} \Sem{P_\Sch}.s_0.(X \ | \ C) \geq \lambda$, for any scheduler $\Sch$. Note also that if C does not hold with a non-zero probability then this definition assumes that the conditional probability is still defined and is maximal.}
\[
\Exp.\Sem{P_\Sch}.s_0.(\Lift{(C \land X)} - \lambda {\times}\Lift{C}) \geq 0 \Wide{\textit{iff}} \Exp.\Sem{P_\Sch}.s_0.(X \ | \ C) \geq \lambda~.
\]
\begin{Prf}
Follows from linearity of the expectation operator and the definition of conditional probability as $\Exp.\Sem{P_\Sch}.s_0.\Lift{(C \land X)}/ \Exp.\Sem{P_\Sch}.s_0.\Lift{C}$ provided that $C$ has a non-zero probability of occurring.
\end{Prf}
\end{Lems}

From \Lem{l1047} we can see that (putting $\lambda=1$) if $\Exp.\Sem{P_\Sch}.s_0.(\Lift{(C \land X)} - \Lift{C}) \geq 0$ then the conditional probability $\Exp.\Sem{P_\Sch}.s_0.(X \ |\ C) = 1$. On the other hand, we can verify the expression  $\Exp.\Sem{P_\Sch}.s_0.(\Lift{(C \land X)} - \Lift{C}) \geq 0$ directly using YAGA's output. Thus the following steps summarise our proposed method for failure mode analysis.

\begin{enumerate}[(a)]
\item Use YAGA to identify a failure tree consisting of traces which terminate in $F$. 

\item From the failure tree identify candidate combinations of events $C$  which correspond to traces terminating in $F$.

\item Using YAGA's output, verify that the candidate combinations $C$ are indeed failure modes by evaluating the constraint $\Exp.\Sem{P_\Sch}.s_0.(\Lift{(C \land X)} - \Lift{C}) \geq 0$ $i.e.$ after setting $\lambda=1$.

\item Compute expected times to failure for the identified failure modes.
\end{enumerate}

In the next section we shall illustrate this technique on a case study of an embedded controller design. 

\begin{figure}
\begin{center}
\includegraphics[scale = 0.17]{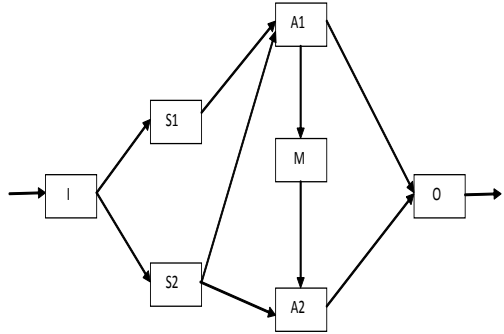}
\end{center}
\caption{An embedded control system.}\label{fig:design}
\end{figure}

\section{Case study two: controller design}\label{cstwo}
Here we show how YAGA can be used to provide important diagnostics feedback to a pB developer summarising the failure the EXPECTATIONS clause in a pB machine refinement. We incorporate the key dimensions of systems dependability --- {\it availability} --- the probability that a system resource(s) can be assessed; {\it reliability} --- the probability that a system meets its stated requirement; {\it safety} --- expresses that nothing bad happens.

The design in \figg{fig:design} is originally based on the work by G\"{u}demann and Ortmeier \cite{GO10}. It consists of two redundant input sensors (S1 and S2) measuring some input signal (I). This signal is then processed in an arithmetic unit to generate the required output signal (O). Two arithmetic units exist, a primary unit (A1) and its backup unit (A2). A1 gets an input signal from both S1 and S2, and A2 only from one of the two sensors. The sensors deliver a signal in finite intervals (but this requirement is not a key design issue since we assume that signals will always be propagated).  If A1 produces no output signal, then a monitoring unit (M) switches to A2 for the generation of the output signal. A2 should only produce outputs when it has been triggered by M. 

An abstract description of the behaviour of the controller is captured in the specification of \figg{fig:SignalTracker}. The reliability of the system is given by the real value $rr$; we encode this in the safety specification within the $expectation(.)$ function. State labels $sg = 2$ and $sg = 3$ denote signal success and failure respectively. Otherwise state labels $sg = 0$ and $sg = 1$ respectively denote idle state and signal in transit.


\begin{figure}
\begin{center}
\scriptsize{
$
\begin{array}{ll}
 \hline\\
\noindent
\mbox{{\bf MACHINE}}  & \mbox{SignalTracker ($maxtime,s1p,s2p,a1p,a2p,mp$)} \\
\mbox{{\bf SEES}} & \mbox{Int\_TYPE, Real\_Type}\\
\mbox{{\bf CONSTRAINTS}} & \mbox{$maxtime \in \mathbb{N} \ \wedge \ s1p,s2p,a1p,a2p,mp \in REAL \ \wedge \ s1p,s2p,a1p,a2p,mp :\in real(0)..real(1)$}\\
\mbox{{\bf CONSTANTS}} & \mbox{$rr$}\\
\mbox{{\bf PROPERTIES}} & \mbox{$rr \in REAL \wedge rr \ge real(0) \wedge rr \le real(1)$}\\
\mbox {{\bf OPERATIONS}  } & \\
\quad \quad \mbox { $sgout \longleftarrow sendsignal \ \Defs \ $} & \\
& \mbox {{\bf PRE} $expectation(real(rr))$ {\bf THEN}}\\
& \quad \mbox{{\bf ANY} $sg$ {\bf WHERE} }\\
& \quad\quad \mbox{$sg \ge 0 \wedge sg \le 3 \wedge expectation(\Lift{(sg = 0 \vee sg = 1)} \times real(rr)  + \Lift{(sg = 2)})$}\\
& \quad \mbox{\bf THEN}\\
& \quad\quad \mbox{$sgout := sg$}\\
& \quad \mbox{\bf END;}\\
\mbox{\bf END}; \\
 \mbox{{\bf END}} & \\
\hline
\end{array}
$
\noindent
}
{\footnotesize}.
\caption{\rm \small Again we use the $expectation(.)$ function to specify that states where $sg = 0 (or 1)$ are worth the system reliability $rr$; states where $sg = 2$ are worth 1 and states where $sg = 3$ are worth 0. This encoding is a safety property for the $sendsignal$ operation and must be preserved by any refinement of the abstract machine.}\label{fig:SignalTracker}
\end{center}
\end{figure}

\subsection{Refining the controller specification}
Here we provide an implementation of the controller by refining the abstract specification in \figg{fig:SignalTracker}.  We also show how to adapt the standard $B$-style modelling of timing constraints \cite{CMR,Butler09} to pB models.  We use the EXPECTATIONS clause of the form $q \Rrightarrow p\times \Lift{(s\neq F)} \Max \Lift{\textit{success}}$, which captures the idea that the probability of reaching the ``success'' state should exceed the given threshold $q$. Here $p$ is a parameter which could vary over the state, but which should initially be at least the value of $q$. Observe that $F$ denotes a state where signal is lost.

But before we do this, we assign individual availability to components of the controller and include the information in the CONSTANTS clause of their abstract machine descriptions. The implementation of the controller as well as the abstract descriptions of its components are in the Appendix. In the next section, we show how to perform dependability analysis on the controller after setting all the components availability to 95\% $(s1p = s2p = a1p = a2p = mp = 0.95)$. To do this, we use YAGA to provide an equivalent MDP interpretation of the refinement in the PRISM language. This then permits experimental analysis of the refinement and hence generation of system diagnostics to summarise the process.

\begin{figure}
\begin{center}
\small{
\begin{verbatim}
    ***** Starting Error Reporting  for Failure Traces located on step 6 *****

                Sequence of operations leading to bad state ::>>>
                [{INIT} (1,0,0,0,0,0), {Sensor2Action} (1,0,1,0,0,0),
            {PrimaryAction} (1,0,1,2,0,0), {MonitorAction} (1,0,1,2,0,2),
    {Skip} (1,0,1,2,0,2), {Sensor1Action} (1,2,1,2,0,2), {SendSignal} (3,2,1,2,0,2)]
                    Probability mass of failure trace is:>>>> 0.00012

                Sequence of operations leading to bad state ::>>>
                [{INIT} (1,0,0,0,0,0), {Sensor2Action} (1,0,2,0,0,0),
            {Sensor1Action} (1,1,2,0,0,0), {PrimaryAction} (1,1,2,2,0,0),
    {MonitorAction} (1,1,2,2,0,2), {Skip} (1,1,2,2,0,2), {SendSignal} (3,1,2,2,0,2)]
                    Probability mass of failure trace is:>>>> 0.00012

                Sequence of operations leading to bad state ::>>>
                 [{INIT} (1,0,0,0,0,0), {Sensor2Action} (1,0,1,0,0,0),
            {PrimaryAction} (1,0,1,2,0,0), {MonitorAction} (1,0,1,2,0,2),
    {Skip} (1,0,1,2,0,2), {Sensor1Action} (1,1,1,2,0,2), {SendSignal} (3,1,1,2,0,2)]
                    Probability mass of failure trace is:>>>> 0.00226

            ************ Finished Error Reporting  ... ***************
\end{verbatim}
}
\end{center}
\caption{ \small Diagnostic feedback revealing single traces at endpoint probability distributions (after setting parameter $maxtime = 6$) corresponding to the failure of the controller to deliver an output signal. Note that the state tuple in this case is given by ($sg$, $s1$, $s2$, $a1$, $a2$,$m$).}\label{fig:diagnostics}
\end{figure}

\subsection{Experiment 1: identification of critical sets}
\noindent{\bf Step 1:}

We set the parameters $q, p := 1$ in  the expression $q \Rrightarrow p \times \Lift{(s\neq F)} \Max \Lift{\textit{success}}$  to identify all failure traces for chosen values of the components availability. \figg{fig:diagnostics} lists three of the failure traces (out of a total of 5) relevant to our discussion, resulting in a maximum probability of failure of 0.0025 after the 6th execution time stamp $i.e.$ $maxtime = 6$.  

\noindent{\bf Step 2:}
From inspection of the above traces we notice that the failure of $A1$ and $M$ enables us to identify them as potential candidates for the construction of our critical set.

\noindent{\bf Step 3:}
We verify that their failure will indeed result in overall failure by examining the value of the expectation $\Lift{(F \land A1 \land M)} - \Lift{(A1\land M)}$.

For candidates such as A1 and M, we use the diagnostic traces to calculate the conditional probabilities as in \Def{d1300}.  To do this we extract all the traces which result in $F$ and then examine the variations of the component failures in the traces to identify those which corresponded to a failure configuration.

 The results were unsurprising and included for example, identifying that a simultaneous failure of the primary unit $A1$ and the backup monitor $M$. On the other hand, once the $pB$ modelling was completed, the generation of the failure traces was automatic improving the confidence of full coverage. To illustrate this point, a programming mistake was uncovered using this analysis where $A1$ was mistakenly programmed to extract a correct reading only if it received signals from both sensors, rather than from at least $1$.

\subsection{Experiment 2: investigating time to failure}

This experiment investigates the time to first occurrence of failure given a particular critical set. In fact, the results show that members of the set of interest are indeed critical after verifying their overall conditional probabilities of failure. In summary, for example, a failure tree corresponding to depth $K=6$ yields distributions over endpoints traces whose components time to failure is shown in \tab{fig:severity}.
\begin{table}
\begin{center}
\small{
\begin{tabular}{|c|c|c|}
 \hline
 \multicolumn{3}{|c|}{Identifying critical components time to first failure} \\
 \hline
Critical Components & Time step to first failure &  Maximum probability of failure \\ \hline
S1, S2  & 2 steps & 2.5000 E-3 \\ \hline
A1, M & 3 steps & 2.4938 E-3 \\ \hline
A1, A2 & 4 steps & 2.4938 E-3 \\ \hline
A1, S2 & 3 steps  & 2.4938 E-3 \\ \hline
\end{tabular}
}
\end{center}
\caption{\rm \small Maximum probabilities of failure are computed with respect to endpoint distributions of failure traces (\figg{fig:diagnostics}) and conditional probabilities are given by \Def{d1300}.}\label{fig:severity}
\end{table}

\section{Related work}

Traditional approaches for safety analysis via model exploration rely on qualitative assessment --- exploring the causal relationship between system subcomponents to determine if some types of failure or accident scenarios are feasible. This is the method largely employed in techniques like the Deductive Cause Consequence Analysis (DCCA) \cite{DCCA}, which provides a generalisation of the Fault Tree Analysis (FTA) \cite{FTA}. Other Industrial methods that support this kind of analysis also include the Failure Modes and Effects Analysis (FMEA) \cite{FMEA} and the Hazard Operability Studies (HAZOP) \cite{HAZOP}. But the efficiency of these techniques is largely dependent on the experience of their practitioners. Moreover, with probabilistic systems, where an interplay of random probabilistic updates and nondeterminism characterise system behaviours, such methods are not likely to scale especially with the dependability analysis of industrial sized systems.

The use of probabilistic model-based analysis to explore dependability features in systems construction has recently become a topical issue \cite{KNP04,pFMEA,GO10,Airbag}. One way to achieve this is to  use probabilistic counterexamples \cite{TJB09,HL09,ADR09} which can guarantee profiles refuting the desired property {\it i.e.} after visiting the reachable states of the supposedly `finite' probabilistic model.

What we have done here is to show how a similar investigation can be achieved for the refinement of proof-based models by taking advantage of the state exploration facility offered by probabilistic model checking. Our method is very precise since it can guarantee the goal of refinement --- improving probabilistic results. However, if this does not hold then we are able to provide exact diagnostics summarising the failure provided that computation resources are not scarce. 

\section{Conclusion and future work}
This paper has summarised an approach based on model exploration for the refinement of proof-based probabilistic systems with respect to quantitative safety specifications in the pB language. Our method can provide a pB designer with information necessary to make judgements relating to dependability features of distributed probabilistic systems. We have shown how this can be done for probabilistic loops hence generalising standard models.

Even though most of the failure analysis conjectured herein have been based on intuition, it should be mentioned that a more interesting investigation would be to explore the use of constraint programming techniques to support full coverage of probabilistic system models. This will enable us target larger refinement frameworks as in \cite{DEPLOY} where probability is not currently being supported.
\\\\
\noindent{\bf Acknowledgement:} The authors are grateful to Thai Son Hoang for assistance with the pB models of the embedded controller. We also appreciate the anonymous reviewers for their very helpful comments.

\bibliography{RefineNet11}
\bibliographystyle{eptcs}

\vspace{15cm}

\begin{figure}
\begin{center}
\scriptsize{
$
\begin{array}{ll}
 \hline
 \fbox{{\bf {\Large Appendix}}} & \\
 \hline\\
\noindent

\mbox{{\bf MACHINE}} & \mbox{Clock ($maxtime$) } \\
\mbox{{\bf CONSTRAINTS}} & \mbox{$maxtime \in \mathbb{N}$}\\
\mbox{{\bf VARIABLES}} & \mbox{$time, action$} \\
\mbox{{\bf INVARIANT}} & \mbox{$time \in \mathbb{N} \wedge action \in \mathbb{N} \wedge time \ge 0 \wedge time \le maxtime$} \\
\mbox{{\bf INITIALISATION}} & \mbox{$time, action := 0, 0$} \\
\mbox {\bf OPERATIONS} &\\
\quad \quad \mbox { $timeout \longleftarrow initClock$ \ $\Defs$ \ } & \mbox {{\bf BEGIN}}\\
& \quad \quad\quad \mbox{$action := 0 \ || \ timeout := 0$}\\
& \mbox{\bf END;} \\
\quad \quad \mbox { $timeout \longleftarrow clockAction (label)$ \ $\Defs$ \ } &\\
& \mbox {{\bf PRE} $label \in \mathbb{N} \wedge time < maxtime$ {\bf THEN}}\\
& \quad \mbox{{\bf BEGIN}}\\
& \quad \quad \mbox{$action := label \ || \ time := time + 1$}\\
& \quad \mbox{\bf END;} \\
&\mbox{\bf END;} \\
&\mbox{$timeout := time$;}\\
\mbox{{\bf END}} &\\\\

\hline
\end{array}
$
\noindent
}
{\footnotesize}.
\caption{\rm \small The specification of the discrete Clock is such that whenever an action due to the components or even a {\bf Skip} action fires, time is incremented while also marking the specific action. We use the action variable as a marker to abstract the identification of the operations constituting the the diagnostic traces (See \figg{fig:diagnostics}). }\label{fig:clock}
\end{center}
\end{figure}

\begin{figure}
\begin{center}
\scriptsize{
$
\begin{array}{ll}
 \hline\\
\noindent
\mbox{{\bf MACHINE}} & \mbox{Cmp ($cp$)} \\
\mbox{{\bf SEES}} & \mbox{Real\_TYPE}\\
\mbox{{\bf CONSTRAINTS}} & \mbox{$cp \in REAL \ \wedge \ cp \ge real(0) \wedge cp \le real(1)$} \\
\mbox {\bf OPERATIONS} &\\\\
\quad \quad \mbox { $cout \longleftarrow componentaction \Defs $} & \mbox{{\bf PCHOICE} \ $cp$ \ {\bf OF} }\\
&  \quad\quad \mbox{$cout := 1$}\\
&  \quad\mbox{{\bf OR}}\\
&  \quad\quad \mbox{$cout := 2$}\\
&  \quad\mbox{{\bf END;}}\\
\mbox{{\bf END}} &\\\\

\hline
\end{array}
$
\noindent
}
{\footnotesize}.
\caption{\rm \small Here we model an abstract stateless machine for components with similar behaviours. Later on, we shall use pB's IMPORT clause to clone Sensor1, Sensor2, PrimaryUnit, Monitor and Backup Units via variable renaming. The specification of the abstract Cmp machine  is such that it can probabilistically either respond to a signal request $(cout = 1 [active])$ or it fails to do so $(cout = 2 [dead])$ . The probability $cp$ is a paremeter of the machine and specifies the availability of the component.}\label{fig:Sensor1}
\end{center}
\end{figure}

\begin{figure}
\begin{center}
\scriptsize{
$
\begin{array}{ll}
 \hline
\noindent

\mbox{{\bf MACHINE}} & \mbox{SignalProcess($s1p,s2p,a1p,a2p,mp$)} \\
\mbox{{\bf CONSTRAINTS}} & \mbox{$s1p,s2p,a1p,a2p,mp \in REAL \ \wedge \ s1p,s2p,a1p,a2p,mp :\in real(0) .. real(1)$}\\

\mbox{{\bf INCLUDES}} & \mbox{Sensor1.Cmp(s1p), Sensor2.Cmp(s2p), PrimaryUnit.Cmp(a1p),}\\
& \mbox{BackupUnit.Cmp(a2p), Monitor.Cmp(mp)}\\
\mbox{{\bf VARIABLES}} & \mbox{$s1,s2,a1,a2,m$} \\
\mbox{{\bf INVARIANT}} & \mbox{$s1, s2, a1, a2, m \in \mathbb{N} \wedge s1, s2, a1, a2, m :: [0,2]$} \\
\mbox{{\bf INITIALISATION}} & \mbox{$s1, s2, a1, a2, m := 0$} \\
\mbox {\bf OPERATIONS} &\\
\quad \quad \mbox { $label \longleftarrow action \Defs$ } &\\
& \quad \mbox {{\bf SELECT} $s1 = 0$ {\bf THEN}}\\
& \quad \quad \mbox {$s1 \longleftarrow Sensor1.componentaction \ || \ label := 1$}\\

& \quad \mbox {{\bf WHEN} $s2 = 0$ {\bf THEN}}\\
& \quad \quad\mbox {$s2 \longleftarrow Sensor2.componentaction \ || \ label := 2$}\\

& \quad \mbox {{\bf WHEN} $a1 = 0 \wedge s1 = 1$ {\bf THEN} }\\
& \quad \quad\mbox {$a1 \longleftarrow PrimaryUnit.componentaction \ || \ label := 3$}\\

& \quad \mbox {{\bf WHEN} $a1 = 0 \wedge s2 = 1$ {\bf THEN} }\\
& \quad \quad\mbox {$a1 \longleftarrow PrimaryUnit.componentaction \ || \ label := 3$}\\

& \quad \mbox {{\bf WHEN} $a1 = 2$ {\bf THEN}}\\
& \quad \quad\mbox {$m \longleftarrow Monitor.componentaction \ || \ label := 4$}\\

& \quad \mbox {{\bf WHEN} $m = 1$ {\bf THEN}}\\
& \quad\quad \mbox {$a2 \longleftarrow BackupUnit.componentaction \ || \ label := 5$}\\

& \quad \mbox {{\bf ELSE} $label := 6$}\\\\

\quad \quad \mbox { $s1out,s2out,a1out,a2out,mout \longleftarrow getState \Defs$} &\mbox{ {\bf BEGIN} s1out,s2out,a1out,a2out,mout := s1,s2,a1,a2,m {\bf END};} \\

\mbox{{\bf END}} &\\

\hline
\end{array}
$
\noindent
}\\
{\footnotesize}.
\caption{\rm \small The nondeterministic behaviour of the components is specified in this machine. An individual component can probabilistically respond to a signal request by setting its state value to 1 or 2 denoting `active' and `dead' respectively, after leaving the initial state with value 0 ('idle').}\label{fig:primaryunit}
\end{center}
\end{figure}

\begin{figure}
\begin{center}
\scriptsize{
$
\begin{array}{ll}
 \hline\\
\noindent
\mbox{{\bf IMPLEMENTATION}} & \mbox{SignalTrackerI($maxtime,s1p,s2p,a1p,a2p,mp$)} \\
\mbox{{\bf REFINES}} & \mbox{SignalTracker} \\
\mbox{{\bf SEES}} & \mbox{Real\_TYPE, Int\_TYPE}\\
\mbox{{\bf IMPORTS}} & \mbox{SignalProcess($s1p,s2p,a1p,a2p,mp$), Clock($maxtime$)}\\
\mbox {\bf OPERATIONS} &\\
\quad \quad \mbox {$sgout \longleftarrow sendsignal $ \ $\Defs$ \ } &  {\bf VAR} \ $sg, s1, s2, a1, a2, m, t$ \ {\bf IN}\\
\quad \quad\quad \quad \mbox{{}} & \mbox{$t \leftarrow initClock$;} \\\\
& \mbox {{\bf WHILE} ($t \le maxtime$) {\bf DO}}\\
& \quad \quad \mbox {$act \longleftarrow action; t \leftarrow clockAction(act);$ }\\
& \quad \quad \mbox {$s1,s2,a1,a2,m \longleftarrow getState$; }\\\\

& \quad \mbox {{\bf IF} $(a2 = 1) \wedge (s2 = 1)$ {\bf THEN} }\\
&  \quad\quad\quad \mbox{$sg := 2$;}\\
& \quad \mbox {{\bf ELSIF} $(a1 = 1) \wedge (s1 = 1)$ {\bf THEN} }\\
&  \quad\quad\quad \mbox{$sg := 2$};\\
& \quad \mbox {{\bf ELSIF} $(a1 = 1) \wedge (s2 = 1)$ {\bf THEN} }\\
&  \quad\quad\quad \mbox{$sg := 2$};\\
& \quad\mbox {{\bf ELSE}}\\
&  \quad\quad\quad \mbox{$sg := 3$};\\
&  \mbox{{\bf END;}}\\
&  \quad\quad\quad \mbox{$sgout := sg$;}\\
& \mbox{{\bf INVARIANT} {\quad $s1, s2, a1, a2, m, t \ \in \mathbb{N} \ \wedge s1, s2, a1, a2, m :: [1,2] \wedge sg :: [0,3] \ \wedge \ t \le maxtime$}}\\
& \mbox{{\bf EXPECTATIONS} \quad $real(rr) \Rrightarrow$ ({\Lift{$(sg = 0 \vee sg = 1)$} $\times real(rr)$  {+ \Lift{$(sg = 2))$}} $\times$ \Lift{$(t = maxtime)$}}}\\
\mbox{{\bf END;}}\\
\mbox{{\bf END}} &\\
\hline
\end{array}
$
\noindent
}
{\footnotesize}.
\caption{\rm \small SignalTrackerI uses a {\bf WHILE-DO} loop structure to model the passage of discrete time. The {\bf PCHOICE} operation provides implementation constructs of the abstract probabilistic branching statements with respect to the availability of the controller components.}\label{fig:implementation}
\end{center}
\end{figure}

\end{document}